\documentclass[sigconf,screen]{acmart}

\AtBeginDocument{%
  \providecommand\BibTeX{{%
    \normalfont B\kern-0.5em{\scshape i\kern-0.25em b}\kern-0.8em\TeX}}}

\usepackage[utf8]{inputenc} % allow utf-8 input
\usepackage[T1]{fontenc}    % use 8-bit T1 fonts
\usepackage{hyperref}       % hyperlinks
\usepackage{url}            % simple URL typesetting
\usepackage{booktabs}       % professional-quality tables
\usepackage{amsfonts}       % blackboard math symbols
\usepackage{nicefrac}       % compact symbols for 1/2, etc.
\usepackage{microtype}      % microtypography
\usepackage{graphicx}
\usepackage{amsmath,amsfonts,amsthm}
\usepackage{array}
\usepackage{subcaption}
\usepackage{bbm}
\usepackage{bm}
\usepackage{multicol}
\usepackage{lipsum}
\usepackage{wrapfig}
\usepackage[ruled]{algorithm2e}
\usepackage{arydshln}
\usepackage{multirow}

\copyrightyear{2022}
\acmYear{2022}
\setcopyright{acmlicensed}\acmConference[WSDM '22]{Proceedings of the Fifteenth ACM International Conference on Web Search and Data Mining}{February 21--25, 2022}{Tempe, AZ, USA}
\acmBooktitle{Proceedings of the Fifteenth ACM International Conference on Web Search and Data Mining (WSDM '22), February 21--25, 2022, Tempe, AZ, USA}
\acmPrice{15.00}
\acmDOI{10.1145/3488560.3501394}
\acmISBN{978-1-4503-9132-0/22/02}

\begin{document}
\fancyhead{}
\title{Tutorial: Modern Theoretical Tools for Understanding and Designing Next-generation Information Retrieval System}


\author{Da Xu}
\affiliation{%
  \institution{Walmart Labs}
  \city{Sunnyvale}
  \state{California}
  \country{USA}
}
\email{Daxu5180@gmail.com}

\author{Chuanwei Ruan}
\affiliation{%
  \institution{Instacart}
  \city{San Francisco}
  \state{California}
  \country{USA}
}
\email{Ruanchuanwei@gmail.com}


% \begin{abstract}

% \end{abstract}






\begin{abstract}
In the relatively short history of machine learning, the subtle balance between engineering and theoretical progress has been proved critical at various stages. The most recent wave of AI has brought to the IR community powerful techniques, particularly for pattern recognition. While many benefits from the burst of ideas as numerous tasks become algorithmically feasible, the balance is tilting toward the application side. The existing theoretical tools in IR can no longer explain, guide, and justify the newly-established methodologies. With no choices, we have to bet our design on black-box mechanisms that we only empirically understand.

The consequences can be suffering: in stark contrast to how the IR industry has envisioned modern AI making life easier, many are experiencing increased confusion and costs in data manipulation, model selection, monitoring, censoring, and decision making. This reality is not surprising: without handy theoretical tools, 
we often lack principled knowledge of the pattern recognition model's expressivity, optimization property, generalization guarantee, and our decision-making process has to rely on over-simplified assumptions and human judgments from time to time. 

Facing all the challenges, we started researching advanced theoretical tools emerging from various domains that can potentially resolve modern IR problems. We encountered many impactful ideas and made several independent publications emphasizing different pieces.
Time is now to bring the community a systematic tutorial on how we successfully adapt those tools and make significant progress in understanding, designing, and eventually productionize impactful IR systems. We emphasize systematicity because IR is a comprehensive discipline that touches upon particular aspects of learning, causal inference analysis, interactive (online) decision-making, etc. It thus requires systematic calibrations to render the actual usefulness of the imported theoretical tools to serve IR problems, as they usually exhibit unique structures and definitions. Therefore, we plan this tutorial to systematically demonstrate our learning and successful experience of using advanced theoretical tools for understanding and designing IR systems. See our webpage for detail: \href{https://moderntoolsfornextgenirs.github.io/}{https://moderntoolsfornextgenirs.github.io/}       

\end{abstract}

\maketitle

\section{Outline}

\begin{figure*}
    \centering
    \includegraphics[width=0.9\textwidth]{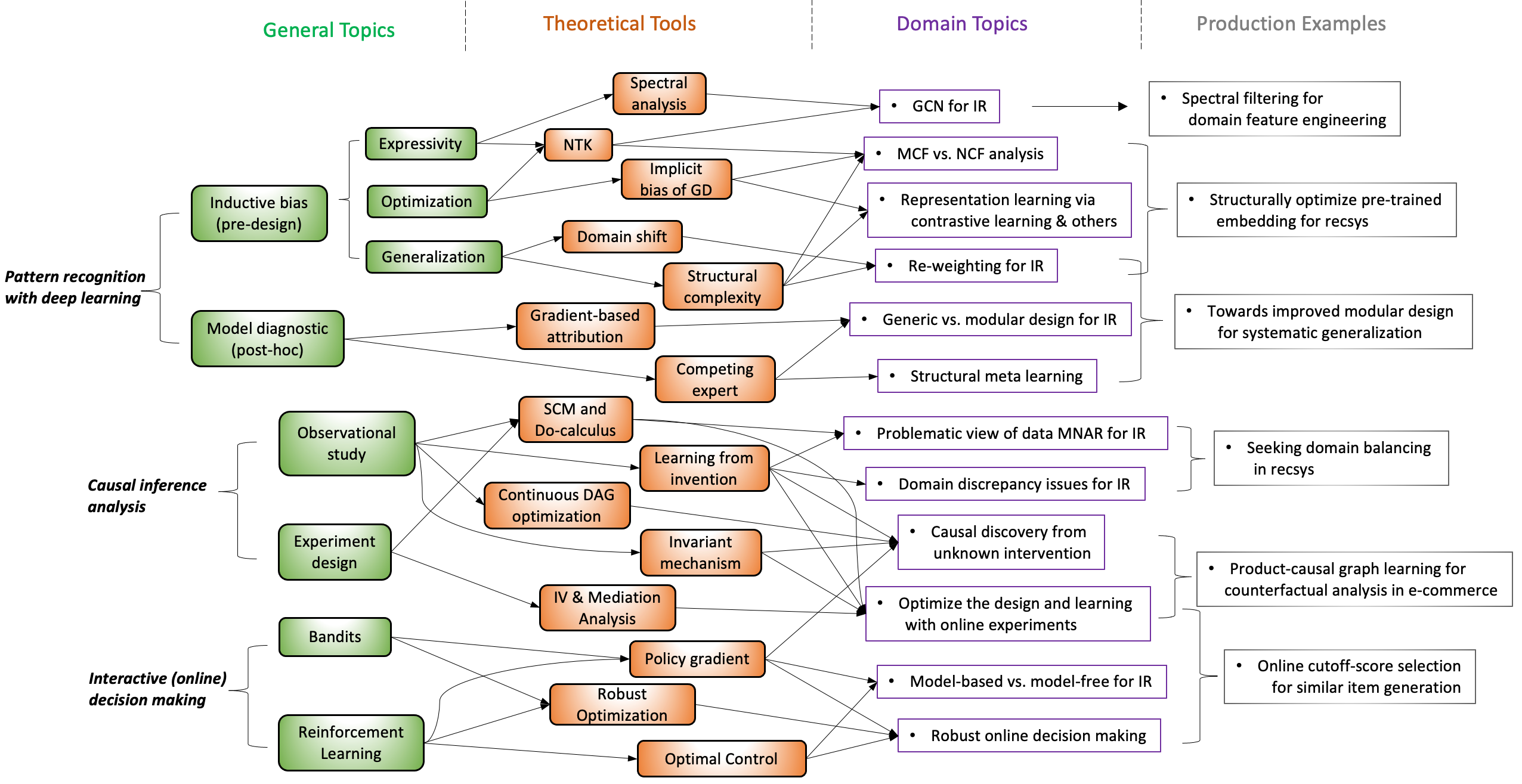}
    \caption{A systematic overview of the (tentative) major topics, theoretical tools, prevalent domain practices, and production examples that will be covered in our tutorial.}
    \label{fig:overview}
    \vspace{-0.15cm}
\end{figure*}

Our tutorial consists of three sections focusing on pattern recognition with deep learning, causal inference analysis, and interactive decision making with bandits and reinforcement learning. We first give an overview of the contents in Figure \ref{fig:overview}, including the major topics, theoretical tools, and their connection with the widespread domain practices and our production examples.

\subsection{Pattern recognition with deep learning}

There are two critical stages for designing and understanding pattern recognition models: the pre-designing stage where we generate a comprehensive inductive bias for the model, and the post-training stage where we diagnose the model to understand why it behaves in certain ways. 
To build up the intuition, image we employ a linear regression model. It immediately becomes clear:
\begin{enumerate}
    \item what the model is \textbf{capable of expressing}: the family of linear functions;
    \item the \textbf{optimization properties}: gradient descent (GD) will lead to the global optimum of convex objectives;
    \item how the model may \textbf{generalize} to unseen data: it will interpolate and extrapolate linearly in the loss-optimal fashion;
    \item why the model would \textbf{perform in particular ways}: the coefficients directly reflect the feature importance.
\end{enumerate}

However, for deep learning models, these questions are incredibly challenging to answer. It causes debates and confusion, such as in \cite{rendle2020neural} where the effectiveness of using neural networks for collaborative filtering (CF) is questioned. In our recent works \cite{xu2021rethinking,xu2021pretrn}, we provide comprehensive and systematic answers to the \emph{expressivity}, \emph{optimization property} and \emph{generalization guarantee} using advanced theoretical tools such as \emph{neural tangent kernel}, \emph{implicit bias of GD} and \emph{structural complexity} of both deep CF models and representation-learning-based methods. We also study out-of-distribution performances of those deep learning IR models by analyzing the impact of \emph{domain shift} \cite{xu2021theoretical,xu2021rethinking}. We also study the widespread domain practice of leveraging graph convolutions networks (GCN) to incorporate the graph-topological signals. Combined with \emph{spectral analysis}, we use the above tools to rigorously show that GCN intrinsically serves as a domain feature engineering approach. As for the final question, we present the powerful tool of \emph{gradient-based attribution} to reveal the impact of each modelling component of complex neural networks. Here, we notice a significant difference between academia and industrial design: researchers tend to develop end-to-end \emph{generic} models that fit a broad range of tasks. Practitioners usually design individual components for particular purposes in the industry and aggregate each tested piece into a larger framework via meta learning. We refer to them as \emph{generic} and \emph{modular} design. We justify and compare them rigorously by combining \emph{gradient-based attribution} with the tool of \emph{competing expert}, which is an extension of the classical \emph{expert voting} method that emphasizes learning. Finally, the instructions for using the mentioned theoretical tools are illustrated via several production examples at \emph{Walmart} and \emph{Instacart}.

\subsection{Causal inference analysis}

The \emph{interventional} nature of IR system is not only reflected in A/B testings, but also the exposures we make (e.g. search result, recommendation, ads displacement) since they can influence users' behavior. According to whether the intervention target is (at least partly) controlled, we have the \emph{experimental} and \emph{observational} settings. The IR community may find the experimental setting more familiar for causal inference analysis, but recently there has been growing interest in using the passively collected observational data for counterfactual reasoning. In both settings, we find Pearl's \emph{do-calculus} and \emph{structural equation} framework \cite{pearl2012calculus} extremely powerful for systematically studying causal problems. For example, we rigorously reveal the fundamental limitations of treating observational feedback with the data missing-not-at-random (MNAR) and the domain-adaptation view. These two domain practices are gaining high popularity, and we also provide promising directions for fixing their issues by discovering \emph{invariant mechanisms} that comprise causality. Towards this end, we introduce advanced optimization and learning tools for \emph{causal discovery} (e.g. continuous directed acyclic graph (DAG) optimization) and \emph{learning from interventional data}. Observational study is not the only focus of our tutorial. We mentioned earlier that existing online experiment frameworks often rely on oversimplified assumptions. By presenting the prominent instrumental variable (IV) and mediation analysis tools, we demonstrate how to lift the assumptions and conduct more robust inferences for IR experiments.

\subsection{Interactive (online) decision making}

Modern IR systems intrinsically build on the understandings of interaction between the information consumer and provider. The \emph{exploration-exploitation} dilemma thus stands out as a major challenge because there are explicit or implicit costs associated with each interaction. Many are motivated to characterize the underlying dynamics using such as \emph{bandits} and \emph{reinforcement learning (RL)}. Setting the practicality issues aside, there are critical conceptual challenges unsettled for both approaches. Notably, there is model-based and model-free options for optimizing the policy in each setting, and while many empirical comparisons have been conducted, there is no rigorous conclusion on what conditions bring the maximum performance out of each candidate. During our investigation, we find the \emph{optimal control} an extraordinary tool and testbed for revealing the strength and weaknesses of many solutions. On the other hand, some empirical studies have found inferior performances from bandits and RL, though they are conceptually more suitable for the tasks. We reveal that this phenomenon is caused by the \emph{robustness} issue: compared with the static counterparts, online decision-making methods are much more sensitive to the algorithmic uncertainty since they tend to accumulate during the process. Therefore, we introduce the advanced theoretical tools from robust optimization and present real-world examples of how to use them to enhance the robustness of our design.

\section{Relevance to WSDM}

WSDM is a prestigious conference hosting the most advanced research work, workshop and tutorial in search and web data mining. Information retrieval, which broadly concerns the process of obtaining the demanded information from a collection of resources produced by information systems, lies at the heart of web search and data mining applications. While developing IR systems is predominantly an engineering effort, the quality of design and the depth of understanding, for both pattern recognition and decision-making procedures, decide the success of any deployment and the sustainability of the development cycle.

As deep learning and other inventions open the door to more complex learning tasks and data-driven decision making, the challenge occurs to the IR community that existing theoretical understandings may not apply to some new technologies. It has thus become a pressing issue to update the methodological tools such that IR enthusiasts can find justifications and support when working on new technologies. 

As researchers and practitioners in the IR community, we have spent considerable efforts in the past few years establishing novel tools, understandings, and theoretical justifications for both pattern recognition and decision making problems, e.g.  \cite{xu2021rethinking,xu2021theoretical,xu2021importance,xu2021towards,xu2020adversarial,xu2021robust}. We also design industrial IR systems powering the online businesses of \emph{Walmart} and \emph{Instacart} in the critical applications of search, recommendation, and advertising \cite{xu2020methods,xu2020product,jie2021bidding,xu2022advances}. Some of the tutorial's content, including arguments, theoretical tools, design ideas, and production examples, are adapted from our previous publications. We further include the cutting-edge results and analytical tools, e.g. \cite{xu21extrapolate,jin2018q,scholkopf2021toward}, classical analysis, e.g. \cite{pearl2012calculus}, as well as novel ideas surveyed from other domains, e.g. \cite{vowels2021d,recht2019tour}, just to list a few.

\section{Format and Schedule}

The tutorial is a half-day event with three sessions:
\begin{enumerate}
    \item pattern recognition with deep learning;
    \item causal inference analysis and experiment design for IR;
    \item interactive (online) decision making with bandits and reinforcement learning.
\end{enumerate}

% A tentative schedule is provided in Table \ref{tab:schedule}.

% \begin{table}[htb]
%     \centering
%     \begin{tabular}{c|l} \hline \hline
%         8:00 - 8:10 AM & Welcome and Opening Remark  \\ \hline
%         8:10 - 9:20 AM & Session on pattern recognition \& case study \\
%         % & (Chuanwei \& Da) \\
%         9:20 - 9:40 AM & Q\&A and coffee break \\ \hline
%         9:40 - 10:20 AM & Session on causal, experiment \& case study \\ 
%         % & (Da) \\
%         10:20 - 10:40 AM & Q\&A and coffee break \\ \hline
%         10:40 - 11:20 AM & Session on decision making \& case study \\
%         % & (Chuanwei \& Da) \\
%         11:20 - 11:40 AM & Q\&A and concluding remark \\ \hline \hline
%     \end{tabular}
%     \caption{Tentative schedule for the half-day tutorial.}
%     \vspace{-0.7cm}
%     \label{tab:schedule}
% \end{table}

% \section{Support Material}

% Since a major part of the tutorial are adapted from our own work, we will be providing the audience with:
% \begin{enumerate}
%     \item A lecture-style writing summary for the detailed contents of our tutorial including all the background;
%     \item Examples of the designing ideas implemented on public dataset (e.g. MovleLens, Amazon Electronics)\footnote{\url{https://github.com/StatsDLMathsRecomSys/WSDM-tutorial}};
%     \item A comprehensive list of resources to the relevant topics;
%     \item The recordings, slides, and research proposals.
% \end{enumerate}

% These material will provide great assistance for audience consuming, revisiting, and building future work (theoretical and applicational) from our tutorial.

\section{Relation with previous tutorial}

We mentioned previously that our tutorial is the first of its kind to the best of our knowledge. We systematically introduce the advanced theoretical tools for understanding and designing modern IR systems. Most existing tutorials focus on the application and engineering aspects of the topics we cover, e.g. \cite{IRBF21,DLIA20,OCE20}. While those tutorial gave detailed descriptions and solutions to specific domain problems, the overlap is minimal since they do not provide comprehensive and systematic introductions to the advanced theoretical tools and connect them to real-world production examples.
In this regard, our tutorial significantly complements those previous efforts by providing self-contained materials that equip both researchers and practitioners with examples, tools, and guidelines to innovate the future of IR with theoretical support.

% \section{Conclusion}

% \newpage

\bibliographystyle{ACM-Reference-Format}
\bibliography{references}


\end{document}